\documentclass[floatfix]{aastex631}
\usepackage{amssymb}
\usepackage{amsmath}
\usepackage{comment}
\usepackage{graphicx}
\usepackage{tabularx}
\usepackage{color}

\usepackage{orcidlink}
\usepackage{lipsum}

\begin{document}

\title{Is the central
	compact object in HESS J1731-347 a hybrid star with a quark core?  An analysis with the constant speed of sound parametrization} 

\author{Suman Pal,\orcidlink{0009-0000-5944-4261}}
\email{sumanvecc@gmail.com}
\affiliation{Physics Group, Variable Energy Cyclotron Centre, 1/AF Bidhan Nagar, Kolkata 700064, India}
\affiliation{Homi Bhabha National Institute, Training School Complex, Anushakti Nagar, Mumbai 400085, India} 
\author{Soumen Podder}
\email{s.podder@vecc.gov.in} 
\affiliation{Physics Group, Variable Energy Cyclotron Centre, 1/AF Bidhan Nagar, Kolkata 700064, India}
\affiliation{Homi Bhabha National Institute, Training School Complex, Anushakti Nagar, Mumbai 400085, India}
\author{Gargi Chaudhuri,\orcidlink{0000-0002-8913-0658}}
\email{gargi@vecc.gov.in}
\affiliation{Physics Group, Variable Energy Cyclotron Centre, 1/AF Bidhan Nagar, Kolkata 700064, India}
\affiliation{Homi Bhabha National Institute, Training School Complex, Anushakti Nagar, Mumbai 400085, India}

\begin{abstract}
In this work, we investigate the possibility of the compact object  in HESS J1731-347 with $M=0.77_{-0.17}^{+0.20}M_{\odot}$ and $R=10.4_{-0.78}^{+0.86} km$ to be a 
hybrid star with quark matter in the inner core.  The observation of this  low mass compact star dictates the use of softer equation of state which on the contrary cannot explain the massive compact stars. This poses a new challenge for the astrophysicists in their search for the equation of state of compact objects. The hybrid equations of state are constructed using the IUFSU parametrization based on the relativistic mean field (RMF) theory for the hadronic  part and the generic constant speed of sound parametrization (CSS) for the phase transition to quark matter.  
 The CSS framework is characterized by three key parameters, namely the transition density $(\rho_{tr})$, the energy jumps $(\Delta\varepsilon)$ and the speed of sound $(C_s)$.  Here, our primary aim is to investigate the influence of individual CSS parameters on  the formation of a object of small mass and radius compatible with HESS J1731-347 parameters. We have also examined the effect of hadronic parameters such as effective mass, symmetry energy, and the slope of the symmetry energy at saturation densities on the formation of this compact object. 
 Finally our analysis suggests that, within a 1$\sigma$ credible level, HESS J1731-347 aligns with the scenario of a stable hybrid star with early deconfinement and higher energy gap .
 
\end{abstract}

\section{Introduction}
The transition between hadronic and quark matter is a highly intriguing area of research with numerous implications in fields such as heavy-ion collision physics, supernova explosions, and binary neutron star mergers (BNSMs). Quantum Chromodynamics (QCD) calculations provide valuable insights into the QCD phase diagram, highlighting the possibility of quark-gluon plasma (QGP) at high temperatures (relevant to the early universe) and at high densities (typical of compact star interiors). While the high-temperature, low-baryon density regime of the QCD phase diagram can be explored through heavy-ion collision experiments, accessing the low-temperature, high-density conditions is much more challenging. These extreme conditions, however, are characteristic of the cores of neutron stars and other compact stars.
A sharp first-order phase transition between hadronic and quark matter can result in a separated branch of hybrid star and neutron star, enabling the formation of twin stars. In this case, two distinct stable configurations can coexist, both with the same mass but differing radii. The larger star will consist solely of hadronic matter, while the smaller, more compact star will be a hybrid star, with a quark core at its center surrounded by hadronic matter. With the advent of binary neutron star merger (BNSM) detections, particularly the observation of GW170817 \cite{LIGOScientific:2018cki}, the phenomenon of phase transitions has attracted significant attention. It has been proposed that if the post-merger phase of the GW170817 event can be observed in the future, analyzing both the inspiral and post-merger stages could offer deeper insights into the properties of compact stars. This analysis may also shed light on the potential for a first-order phase transition occurring during the merger process.

In a recent study, \cite{2022NatAs} utilized modeling of the X-ray spectrum and the Gaia-observation-based distance to estimate the mass  and radius  of the central compact object (CCO) in the supernova remnant HESS J1731-347. Their best-fit values, determined at the 1$\sigma$ posterior credible levels, are $M=0.77_{-0.17}^{+0.20} M_{\odot}$ and  $R=10.4_{-0.78}^{+0.86} km$, respectively.
This estimation identifies the CCO in HESS J1731-347 \cite{2022NatAs} as the lightest neutron star discovered so far and possibly a candidate for an exotic object, such as a strange quark star \cite{Horvath:2023uwl,Gholami:2024ety,Rather:2023tly}.
It is important to acknowledge that, although considerable efforts have been made to analyze HESS J1731-347, exploring its nature as either a conventional neutron star or a hybrid star, its extraordinary properties remain a topic of ongoing discussion and scrutiny. In the event of the  observational constraints having reached new horizons at recent times,  the study becomes more challenging with respect to the constituents of the compact stars as well as the possibility of hadron-quark phase transition at its core. Most interestingly, recently observed HESS J1731-347 \cite{2022NatAs} remnant suggests a very small mass while PSRJ0740+6620 \cite{Fonseca:2021wxt} suggests a mass as high as $ M=2.08\pm 0.07  M_{\odot}$. Recently, \cite{Salmi:2022cgy,Salmi:2024aum,Dittmann:2024mbo} re-analysed the radius of the PSR J0740+6620 to be slightly increased $R=12.49_{-0.088}^{+1.28}$ km.  from \cite{Miller:2021qha,Riley:2021pdl}.
Moreover, PSR J0030+451 is further illuminated by data from NICER's measurements \cite{Riley:2019yda, Miller:2019cac} and has also recently been reported by the NICER Collaboration \cite{Choudhury:2024xbk,Reardon:2024rdv}.

This allows for a wide range of equations of state
with varying stiffness. Also, several works \cite{prd_Monta_2019_may, PhysRevD.105.023018} have explored that the gravitational wave data is consistent with the occurrence of twin stars. The detection
of gravitational wave (GW170817 \cite{LIGOScientific:2018cki}) from binary NS merger
(BNSM) provides additional constraints on the equation of
state (EOS).

The recent discovery of this low mass as well as low radii compact object has presented a lot of challenges to the astrophysics community regarding the equation of state of this interesting object. There has been a
lot of speculation \cite{Sagun:2023rzp,Mariani:2024gqi,Char:2024kgo,Veselsky:2024eae,Tewari:2024qit,Kourmpetis:2024mol,Zhang:2024ldq,Laskos-Patkos:2023tlr} regarding the composition of the compact object in HESS J1731-347  ranging from the lightest observed neutron star \cite{Kubis:2023gxa}, a strange quark star \cite{Horvath:2023uwl,Rather:2023tly,Gholami:2024ety}, a hybrid star with an early deconfinement phase transition, or a dark matter–admixed neutron star \cite{Hong:2024sey,Pal:2024afl}. Recently there has been  few studies considering this to be a hybrid star \cite{Laskos-Patkos:2023tlr,Mariani:2024gqi}.
Researchers have extensively employed different phenomenological quark models like the MIT bag model \cite{chodes1974, Sen:2021cgl,Sen:2022lig, Pal:2023dlv, Podder:2023dey}, quark mass model\cite{peng2001a,wen2005a, Chu_2014, Benvenuto95} or quasi particle model \cite{Zhang:2021qhl,pen:23prc_qmdd, Ma:prd23sep}, NJL model \cite{nambu1961njl,Klevansky92,Hatsuda:1994pi,Buballa:2003qv,Buballa:1998pr,Lenzi:2010mz,Hanauske:2001nc,Wang:2020wzs,Pfaff:2021kse}, the perturbation model \cite{Fraga2001}, the field correlator method \cite{Plumari2013}, the quark-cluster model \cite{Xu2003} and many other models to investigate the thermodynamic properties of strange quark matter, quark stars, and hybrid stars. These models typically account for all interactions among quarks through bag pressure or an equivalent quark mass. However, it is relatively simpler and at the same time justified to deal with a constant speed of sound model(CSS) \cite{PhysRevD.88.083013,Tsaloukidis:2022rus,Laskos-Patkos:2023cts} for exploring hybrid stars.

The investigation of phase transition in hybrid stars (HSs) has produced several intriguing findings and opportunities like exploring the speed of sounds($C_s$) in dense matter.  The bare minimum that the equations of state must meet is that $0\le C_s^2\le1$, as suggested by causality and thermodynamic stability. The possibility of upper bounds on the velocity of sounds($C_s^2$) in a dense environment is more interesting and challenging. Zel’dovich \cite{Zeldovich:1961sbr,zel2014stars} first pointed out with the help of covariant formulations that $C_s^2\le\frac{1}{3}$. In calculations using holography \cite{PhysRevD.80.066003} and lattice QCD \cite{BORSANYI201499}, $C_s^2$ is found to be less than $\frac{1}{3}$ at high temperatures and zero baryon chemical potential. But at low temperatures and high baryon density, the situation is quite different. In recent studies on hybrid stars \cite{PhysRevC.102.055801,suman_2023_prd1} it was proposed that
$C_s^2$ must be greater than $\frac{1}{3}$ in order to satisfy the recent astrophysical constraints($M\ge2M_{\odot}$) and also in \cite{Altiparmak_2022} it has been shown using a large number EoSs and statistical distributions that it is natural to expect that $C_s^2$ is greater than $\frac{1}{3}$.

The possibility of exploring the effect of varying  different parameters in CSS model makes it a preferable one for the analysis of the phase transition in the hybrid stars. 
We examine the properties of the hybrid stars with the help of the speed of sound, the energy gap and the transition density and therefore explore the effect of these parameters in satisfying the criteria of mass and radius of the small compact object (HESS J1731-347). The investigations of the hybrid stars can give rise to a new stable branch of so-called twin stars \cite{SCHERTLER2000463}. This work will demonstrate that the star on the 2nd branch could be the small object in HESS J1731-347, having a twin in the hadron branch. The primary aim of this study is to investigate the nature of the CCO in the HESS J1731-347 remnant within the frame-
work of the hybrid star model using the constant speed of
sound approach.
 In addition to this, we vary the parameters of the RMF model of the hadron sector in order to study their effect on compact object (HESS J1731-347) properties.

This paper is organized as follows. In Sec.\ref{Formalism}, we will briefly outline the formalism of the model used for hadronic and quark sectors. In Sec.\ref{Results} we present our results on the possible configurations of HESS J1731-347, considering different parameters involved in the quark sector as well as the hadron sector. In Sec.\ref{sec:Variation of the parameters of the hadronic EoS} we examined the impact of variation of the parameters of the hadronic EoS on hybrid star configurations.  Finally, we summarize our results in Sec.\ref{sec:conclusion}

\section{Formalism}
\label{Formalism}
\subsection{Pure hadronic phase}
\label{HP} 
 We consider the relativistic mean field (RMF) model for the description of the hadronic part in HSs. We consider the cold-dense matter equation of state which is a key component of the compact star equation of state. The detailed description of the RMF model is given in \cite{Glendenning:1997wn}. We adopt the widely used IUFSU parametrizations \cite{Piekarewicz_prc_2010_nov} and consider $\beta$ equilibrated and charge neutral matter consisting of n, p, $e$, and $\mu$. In this work, the presence of higher degrees of freedom like the hyperons are not considered. The saturation properties at saturation density($\rho_0$) namely the binding energy per nucleon ($\frac{BE}{A}$), nuclear incompressibility ($\text{K}_{\text{sat}}$), symmetry energy coefficient ($\text{E}_{\text{sym}}$) and symmetry energy slope parameter ($\text{L}_{\text{sym}}$) of the models are given in Table[\ref{tab:1}]. The chosen $\text{E}_{\text{sym}}$ and $\text{L}_{\text{sym}}$ of the hadronic matter are consistent with the recent findings obtained from the PREX-II \cite{Adhikari_PRL_2021} experiment. 
 The detailed calculation of the algebraic determination of the coupling constants from the nuclear matter saturation properties is shown in \cite{Glendenning:1997wn,Hornick:2018kfi,Chen:2014sca}.
 
\begin{table}[!ht]
\caption{The nuclear matter properties at saturation density $\rho_{_{0}}$ for the hadronic model(IUFSU\cite{Piekarewicz_prc_2010_nov}).}
\setlength{\tabcolsep}{20.0pt}
\begin{tabular}{cccccc}
\hline
\hline
 $\rho_{0}$ & $B/A$ & $K_{sat}$ &$\text{m}^*/\text{m}$  & $\text{E}_{\text{sym}}$ &$\text{L}_{\text{sym}}$\\
 $({\rm fm}^{-3})$ & (MeV) & (MeV) &  &(MeV) & (MeV)\\ \hline
 0.155 & $-$16.40 & 231.2 & 0.61& 31.30 &47.2\\
\hline
\hline
\end{tabular}
\label{tab:1}
\end{table}
\subsection{Hadron-quark phase transition} 
\label{PT} 

The transition from the hadronic phase to the quark phase, which occurs as a first-order phase transition, can be described using either the Gibbs construction (GC) or the Maxwell construction (MC), depending on the value of the surface tension at the hadron-quark boundary \cite{Maruyama:2007ss,Maruyama:2007ey}. The GC relies on the global charge neutrality condition, leading to the formation of a mixed phase, whereas the MC enforces local charge neutrality, resulting in a significant density discontinuity. In the MC approach, the transition occurs at constant pressure, leading to a sharp phase transition, whereas in the GC approach, the transition is smoother with a gradual pressure variation. 
Therefore, the equation of state exhibits significant differences in the mixed-phase region. In the GC approach, the transition begins at a lower density and extends over a broader range of energy densities, leading to a gradual softening of the equation of state and consequently affecting the mass-radius relation. For more details, one can refer to \cite{Montana:2018bkb,Pereira:2022stw,Sun:2023glq}.
Another important aspect is the speed of sound in quark matter, as mentioned in the introduction. 
One can also utilize the density-dependent speed of sound parametrization details provided in \cite{Tews:2018kmu, Margaritis:2019hfq}. It has been demonstrated in \cite{Margaritis:2019hfq} that the upper bound on the speed of sound ($C_s^2 = 1$) in dense matter significantly influences the bulk properties of maximally rotating neutron stars. Moreover, the imposition of a lower bound ($C_s^2 = \frac{1}{3}$) results in a substantial reduction in the maximum mass of these stars.
For a detailed study of the realistic speed of sound in the MIT bag model (with various forms), the quasi-particle model, and the NJL model, one can refer to \cite{Pal:2023quk,Pal:2024nza,Ranea-Sandoval:2015ldr,Han:2019bub}. 
The introduction of the repulsive interaction in these realistic models increases the value of the speed of sound from the conformal limit( $C_s^2 = \frac{1}{3}$), thereby increasing  the stiffness of the equation of state, aligning it with recent astrophysical observations, especially the $2\,M_{\odot}$ constraint.

\color{black}
In this work, we consider the hybrid star equation of state and we employ the simple CSS parametrization to describe the hadron-quark phase transition and the quark core. The CSS model consists of three free parameters: transition density ($\rho_{tr}$), energy density jump ($\Delta\varepsilon$), and square of the speed of sound ($C_s^2$). 
For the phase transition part, we consider the standard MC  technique where a sharp hadron to quark phase transition occurs at the transition density. We have adopted the parametrization introduced by \cite{PhysRevD.88.083013,Tsaloukidis:2022rus,Laskos-Patkos:2023cts}. Below the transition pressure($P_{tr}$) or transition density ($\rho_{tr}$), we consider hadronic EoS($\varepsilon_H$)  whereas above the transition point, we use the constant speed of sound parametrizations for the quark matter. The energy density below and above the transition density is as follows:
\begin{equation}
    \varepsilon(P)=\begin{cases}
    \varepsilon_H, & \text{if } P \leq P_{tr} \\
    \varepsilon(P_{tr}) + \Delta \varepsilon + \frac{1}{C_s^2}(P-P_{tr}), & \text{if } P > P_{tr}
    \end{cases}
\end{equation}

We compute hybrid EoS with the different parameters mentioned above. To construct the 
crust, we have used Baym, Pethick, and Sutherland (BPS)\cite{BPS_1971} EoS for the outer crust and for the inner crust, we have employed the polytropic form of the EoS as given in \cite{Carriere:2002bx}.  To study the structural properties of the hybrid stars under static conditions, such as their gravitational mass (M) and radius (R), we use the Tolman-Oppenheimer-Volkoff (TOV) \cite{PhysRev.55.364, PhysRev.55.374} equations based on the hydrostatic equilibrium between gravity and internal star pressure. 
\begin{figure*}[htp]
    \centering
      \includegraphics[width=0.45\textwidth]{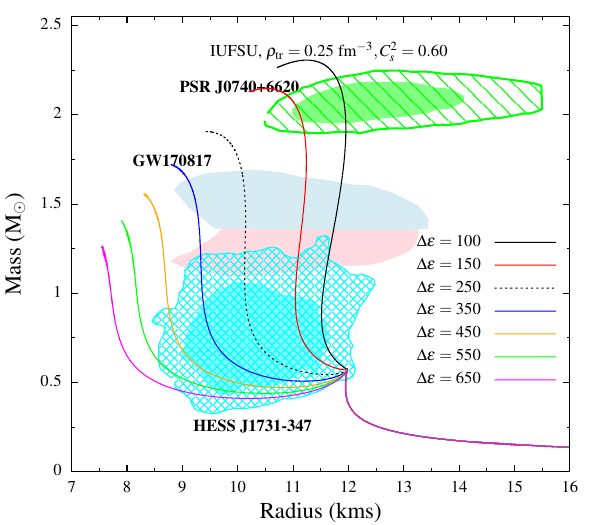}
      \includegraphics[width=0.45\textwidth]{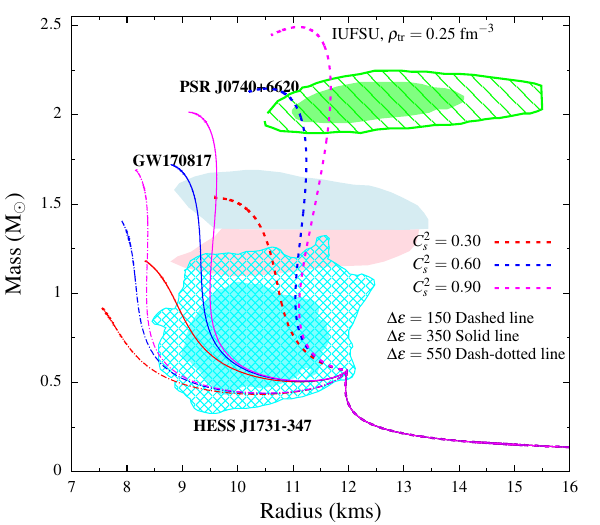}
\caption{\it Mass-radius relationship of hybrid star configuration with IUFSU and CSS parametrization; plots for a fixed $\rho_{\text{tr}}=0.25 ~\text{fm}^{-3}$ for different values of $\Delta \varepsilon$($MeV fm^{-3}$) for a fixed $C_s^2$ (left panel) and with different values $C_s^2$ and $\Delta \varepsilon$ (right panel). Observational limits imposed from PSR J0740+6620 \cite{Fonseca:2021wxt,Miller:2021qha,Salmi:2022cgy,Salmi:2024aum} and  HESS J1731-347 \cite{2022NatAs}  indicated. The constraints on the $M-R$ plane prescribed from GW170817 \cite{LIGOScientific:2018cki}  are also compared.}
\label{fig:mr_0.25}
\end{figure*} 

\begin{figure*}[htp]
    \centering
      \includegraphics[width=0.45\textwidth]{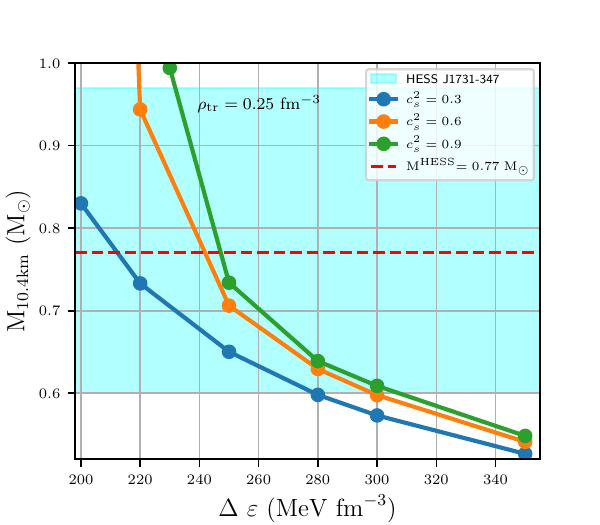}
      \includegraphics[width=0.45\textwidth]{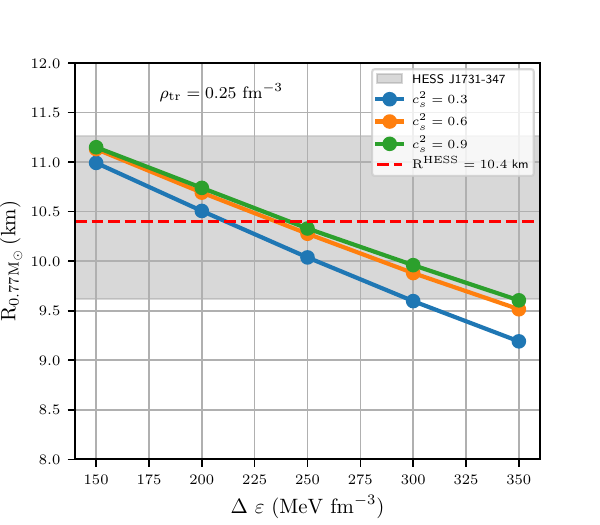}
      \caption{  The left panel shows the variation of $M_{\text{10.4km}}$ (mass of stars with radius =10.4km) with $\Delta \varepsilon$ for different speeds of sound. The observed mass range ($M=0.77_{-0.17}^{+0.20}M_{\odot}$) for the CCO in HESS J1731-347 has been shaded with cyan color. The right panel illustrates the variation of $R_{0.77 M_{\odot}}$ ((radii of stars with mass=0.77$M_{\odot}$)) with $\Delta \varepsilon$ for different speeds of sound. The observed radii range ($R=10.4_{-0.78}^{+0.86} km$) has been shaded with grey. The transition density ($\rho_{\text{tr}}$)  is 0.25~$\text{fm}^{-3}$. }
\label{fig:mr_0.25_1sigma}
\end{figure*}

\begin{figure*}[htp]
    \centering
      \includegraphics[width=0.45\textwidth]{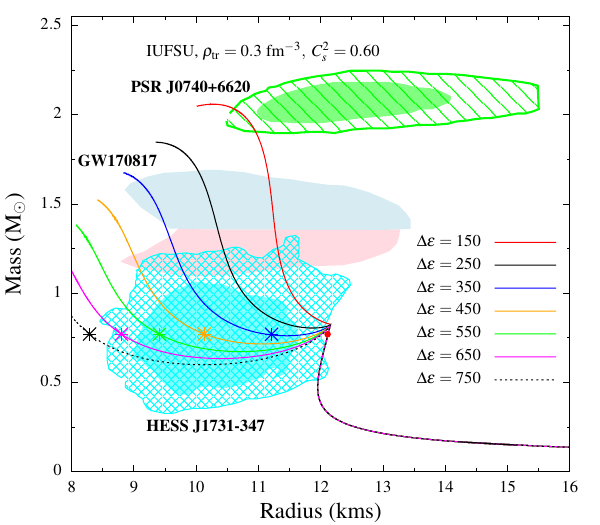}
     \includegraphics[width=0.45\textwidth]{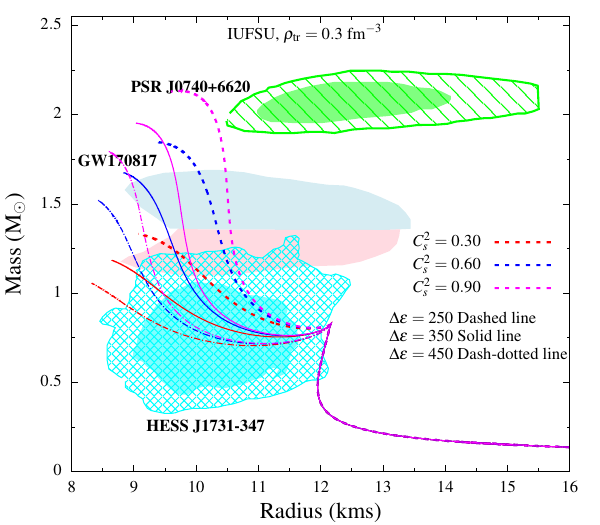}
\caption{ Same as Fig.~\ref{fig:mr_0.25} with $\rho_{\text{tr}}=0.30 ~\text{fm}^{-3}$. In the left panel, the solid red points represent the mass= 0.77 $M_{\odot}$ in the hadronic branch, and the star marks with different colors represent the corresponding twin stars in the hybrid branch}
\label{fig:mr_0.30}
\end{figure*} 

\begin{figure*}[htp]
    \centering
      \includegraphics[width=0.45\textwidth]{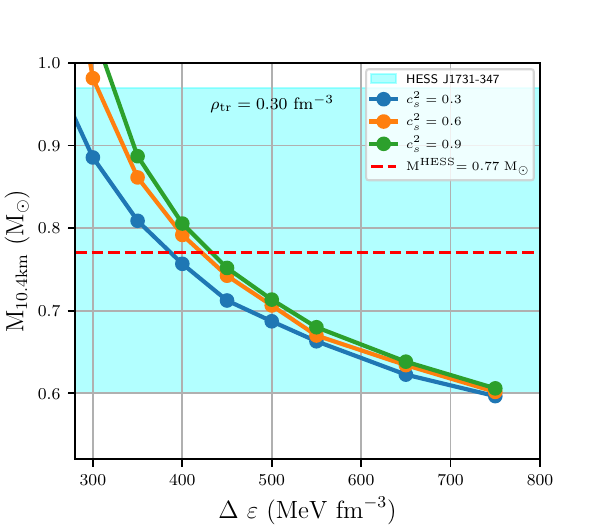}
       \includegraphics[width=0.45\textwidth]{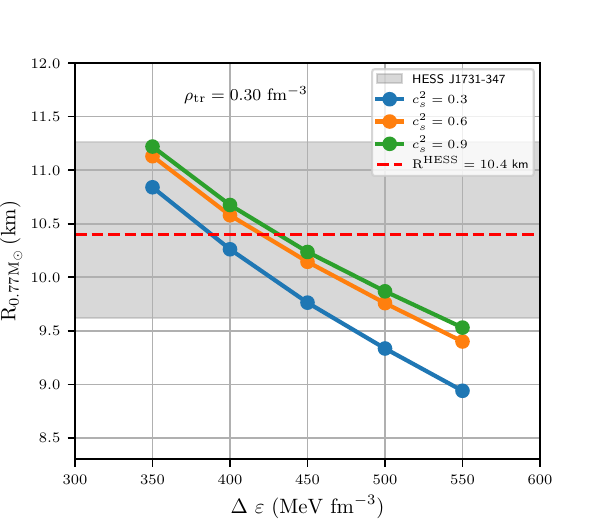}
       \includegraphics[width=0.45\textwidth]{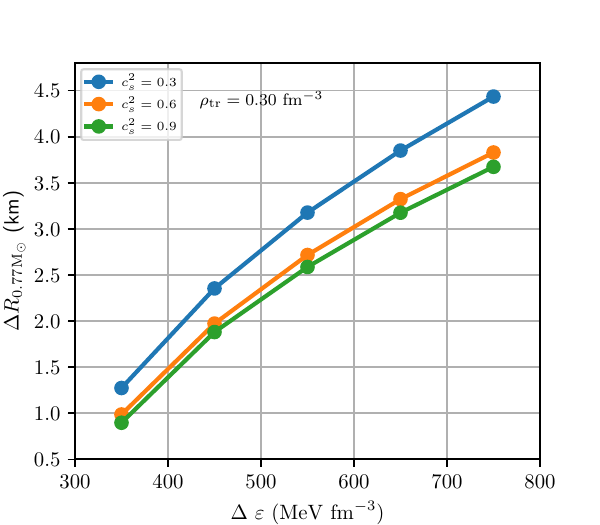}
      \caption{ The top left panel figure is same as Fig. 2(left panel) with $\rho_{tr} =0.3~\text{fm}^{-3}$. The top right panel figure is same as Fig. 2 (right panel) for $\rho_{tr} =0.3~\text{fm}^{-3}$ . The bottom panel shows the twin-star radii difference for $M = 0.77~M_{\odot}$ vs. $\Delta \varepsilon$ for various $C_s^2$ with $\rho_{tr} = 0.3~\text{fm}^{-3}$.}
\label{fig:mr_0.30_1sigma}
\end{figure*}
\begin{figure*}[htp]
    \centering
         \includegraphics[width=0.45\textwidth]{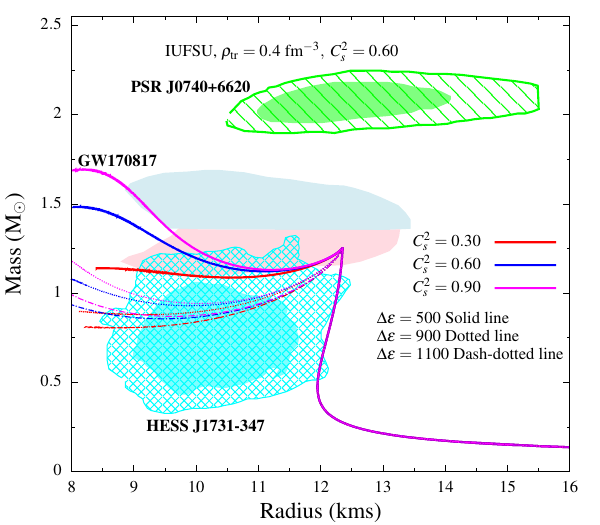}
     \includegraphics[width=0.45\textwidth]{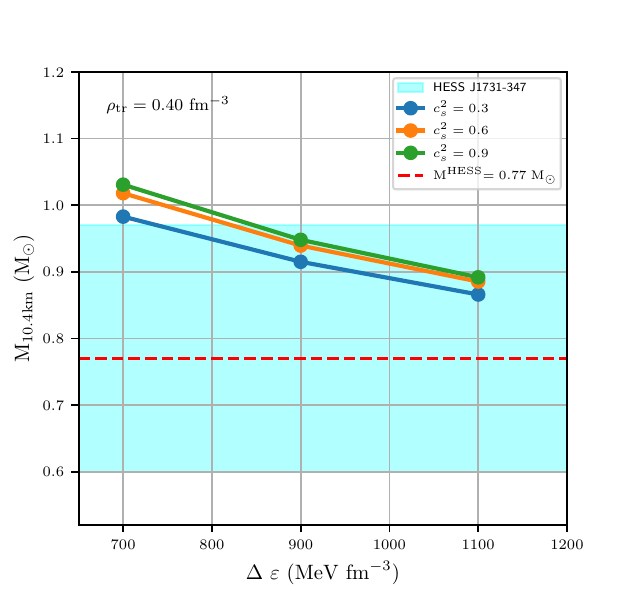}
\caption{\it  The left panel shows the mass-radius relationship for the different values of $C_s^2$ along with different values of $\Delta \varepsilon$ for the transition density ($\rho_{\text{tr}}$)=0.40 $\text{fm}^{-3}$.
The right panel  is same as Fig. 2 (left panel) for $\rho_{\text{tr}}$=0.40 $\text{fm}^{-3}$.}
\label{fig:mr_0.40}
\end{figure*}

\begin{table}[!ht]
\caption{ Energy density, pressure, mass, and radius for the hadronic sector at the transition densities. }
\setlength{\tabcolsep}{20.0pt}
\begin{tabular}{cccccc}
\hline
\hline
 $\rho_{tr}$ & $ \varepsilon_{tr}  $ & $ P_{tr}  $ & $\Delta \varepsilon_{cr}$ &$ M_{tr}  $  & $R_{tr}$ \\
 \hline 
  $\text{fm}^{-3}$ & MeV $\text{fm}^{-3} $ &  MeV $\text{fm}^{-3} $  & MeV $\text{fm}^{-3} $ & $M_{\odot}$ & km \\
 \hline   
0.25 & 241.86 & 11.65 &138.40 &0.5709 & 11.976 \\
0.30 & 293.46& 20.48& 177.45 & 0.8247& 12.156 \\
0.40 & 402.05 & 46.23 & 270.37& 1.2568 &12.349 \\
\hline
\hline
\end{tabular}
\label{tab:2}
\end{table}

\section{Result and Discussions}
\label{Results}
The  EoSs are constructed via IUFSU \cite{Piekarewicz_prc_2010_nov} and CSS parametrizations \cite{PhysRevD.88.083013} which describe the hadronic matter and the quark matter respectively. The phase transition has been executed through Maxwell Construction (MC).
In this work, we calculate the structural properties which include radius (R) and gravitational mass (M) of the compact objects. The possible existence of the compact star with a mass less than 1$M_{\odot}$ has been examined w.r.t different parameters like $\rho_{tr}$, $\Delta \varepsilon$ and $C_s^2$ of the CSS parametrization. The transition values of energy density ($\varepsilon_{tr}$), pressure ($P_{tr}$), mass ($M_{tr}$) and radius ($R_{tr}$) corresponding to the three chosen transition densities ($\rho_{tr}$) at the point of quark hadron phase transition are displayed in Table \ref{tab:2}.

To begin with, we consider a fixed transition density of 0.25 $\text{fm}^{-3}$ in the hadronic sector and fixed values of the speed of sound ($C_s^2$= 0.60) in the quark sector with different energy density gap ($\Delta \varepsilon$) as shown in the left panel of  Fig.~\ref{fig:mr_0.25}. The compact objects in the hybrid branch satisfy the mass and radii constraints of the remnant in HESS J1731-347  at 1$\sigma$ level (parallax priors and X-ray data) for all values of the energy density gap parameter used starting from 100 MeV $\text{fm}^{-3} $. The results using the gap parameter in the range of (100-350) MeV $\text{fm}^{-3} $ satisfy well the GW170817  data while the lowest values of the energy gap 100 and 150 MeV $\text{fm}^{-3} $ satisfy well the constraints from PSR J0740+6620, the maximum mass reaching 2 solar mass.   In the right figure we study the Mass-Radius characteristics for the same  fixed  transition density, varying the constant speed of sound parameter as well as the energy gap. It has been observed that all the EoS for this case satisfy the constraints from the remnant in HESS J1731-347 irrespective of the speed of sound for the energy gaps  used.  The low radii of the compact object is better satisfied by the EoS having higher energy gaps. The maximum mass in the hybrid branch increases when speed of sound is increased. The results from the energy gap of 150 MeV $\text{fm}^{-3} $ satisfy the constraints from GW170817 reasonably well.  The maximum mass reaches the 2 solar mass constraint for $C_s^2$=0.6 and 0.9 with $\Delta \varepsilon=150$ Mev $\text{fm}^{-3}$

In the next part (Fig.\ref{fig:mr_0.25_1sigma}), we intend to show the compatibility of the properties of the  compact object with the results obtained from our hybrid equations of state by showing directly the variation of mass and radii with that of the energy gap for different speeds of sound. In order to better demonstrate that,  we have plotted the masses ($M_{10.4km}$) of the compact objects formed in the stable hybrid branch from our sets of  EoS having radius =10.4 km (as estimated for the remnant in HESS J1731-347  \cite
{2022NatAs}). Results have been demonstrated as a function of different gap parameter $\Delta \varepsilon $ for  three values of the speed of sound. The allowed mass range ($M=0.77_{-0.17}^{+0.20}M_{\odot}$) for this compact object has been shaded with cyan. The masses ($M_{10.4km}$) lie within the range estimated in Fig. \ref{fig:mr_0.25_1sigma}  for the range of of $\Delta \varepsilon $  nearly from 200 to 300 MeV $\text{fm}^{-3} $. Both lower and higher values of $\Delta \varepsilon$ which satisfies the mass constraints as seen from the Fig.~\ref{fig:mr_0.25_1sigma}(left panel) increases  with the increase in the speed of sound .
The change in the mass as speed of sound is varied from 0.3 to 0.9 is not much for higher energy gaps;  the mass changes more significantly as $\Delta \varepsilon$  is being changed.  

In the right panel of the Fig.~\ref{fig:mr_0.25_1sigma}, we plot the radii   $R_{0.77 M_{\odot}}$obtained for different compact stars corresponding to the mass 0.77 solar mass as estimated for the  compact object in the remnant of HESS J1731-347. The observed radii range ($R=10.4_{-0.78}^{+0.86} km$) for this compact object has been shaded with grey. The radii has been plotted as function of the energy gap parameter for different $C_s^2$ values. The radii decreases as the energy gap increases , while for a fixed energy gap it decreases as speed of sound is reduced.  As the energy gap increases beyond 300 MeV $\text{fm}^{-3} $, the radii as found from our calculations goes beyond the observed range. These two figures  also show that the constraints from this compact object is less sensitive to the velocity of sound used.

In the next phase of our calculation, we increase the  value of transition density($\rho_{tr})$  to 0.3 $\text{fm}^{-3}$ in the hadronic sector keeping  $C_s^2=0.60$ in the quark sector and consider different values of energy gap $(\Delta \varepsilon)$ as shown in the left panel in Fig.~\ref{fig:mr_0.30}. HESS J1731-347 \cite{2022NatAs} remnant constraint  with small mass and radius is satisfied with all values of $\Delta\varepsilon$ (250-750 MeV $\text{fm}^{-3} $) at 1$\sigma$ (parallax priors and X-ray data)  while the lowest value of 150 MeV $\text{fm}^{-3} $ is satisfied at 2$\sigma$ (parallax priors and X-ray data)for this transition density. 
The compact stars in the stable branch after phase transition satisfies well the constraints of mass and radii from the remnant in HESS J1731-347. 
{
The criterion for a first-order phase transition in neutron stars to generate an unstable region was considered by Seidov  \cite{seidov1971stability,Laskos-Patkos:2023cts}. The existence of a third family of compact stars requires the appearance of an unstable region in the mass-radius diagram, followed by a stable one. The minimum jump in energy density required for the emergence of an unstable configuration is determined by the following relation:
$ \Delta \varepsilon_{cr}=\frac{1}{2} \varepsilon_{tr}+ \frac{3}{2} P_{tr} $.
In this work, we consider three transition densities, with their respective critical energy jumps listed in Table~\ref{tab:2}. Based on the values of $\Delta \varepsilon_{cr}$ and $\Delta \varepsilon$, hybrid stars can be categorized into four distinct classes as mentioned by ~\cite{PhysRevD.88.083013}. Since we are interested in describing HESS J1731-347 as a twin star with a mass of $0.77 M_{\odot}$, a transition density of $\rho_{tr} = 0.30~\text{fm}^{-3}$ is found to be appropriate. 
For $\rho_{tr} = 0.30~\text{fm}^{-3}$, the critical energy gap is $\Delta \varepsilon_{cr} = 270.37$ MeV $\text{fm}^{-3} $, and a HESS J1731-347  twin star configuration is obtained for an energy gap $\Delta \varepsilon$ greater than this critical value. According to \cite{PhysRevD.88.083013}, this results in a disconnected hybrid branch.

}
These compact objects are mostly part of the twin stars, the other component  being located in the hadronic branch which satisfies the constraints at 2$\sigma$ level. The maximum mass of the twin stars is dictated by the maximum mass in the hadronic branch.{\color{blue}} For $\rho_{tr}$ = 0.3 $\text{fm}^{-3}$, this maximum mass is roughly equal to 0.82 $M_{\odot}$ and the radii of this is approximately 12.75 km in the hadronic sector. The compact objects in the hybrid branch  as part of the twins are of lower radii ( marked by stars in the figure) compatible with the HESS J1731-347 observation. The maximum mass of the stars in this stable hybrid branch satisfies the constraints from different observation for different values of the energy gap. The one with the lowest energy gap (150 MeV $\text{fm}^{-3} $) satisfies the constraint of 2 $M_{\odot}$ from PSR J0740+6620.  The results with  values of the energy gap from 150 to 350 MeV $\text{fm}^{-3} $ satisfies the data from GW170817.   It
is quite interesting  that at this transition density,  this parameter set with varying energy gap satisfies 
the lower mass (HESS J1731-347 \cite{2022NatAs}) as well as the higher
mass compact stars (PSR J07040+6620 \cite{Fonseca:2021wxt,Salmi:2022cgy,Salmi:2024aum} ) from the hybrid
star scenario.  
In the right panel of Fig.~\ref{fig:mr_0.30}, we study the Mass-Radius characteristics for the same fixed transition density, varying the constant speed of sound parameter and the energy gap, similar to the case of the transition density 0.25 $\text{fm}^{-3}$.

Next in Fig.~\ref{fig:mr_0.30_1sigma}, we repeat the plots of Fig.~\ref{fig:mr_0.25_1sigma} for the  transition density 0.30 $\text{fm}^{-3}$. The observational constraints on the mass restricts the  range energy gap ($\Delta \varepsilon$) to nearly 450 MeV $\text{fm}^{-3} $ (300 to 750 MeV $\text{fm}^{-3} $) at this density as observed from the Fig.~\ref{fig:mr_0.30_1sigma}. The adjoining plot (right panel) of the radii corresponding to the mass 0.77 $M_{\odot}$ restricts the energy gap to a  smaller interval of less than 200 MeV $\text{fm}^{-3} $ (350 to 525 MeV $\text{fm}^{-3} $). 
From these two figures, it is evident that the limits of the radius of the compact object as estimated from the remnant put more constraints on the equation of state ($\Delta \varepsilon$).

In the bottom panel of Fig.~\ref{fig:mr_0.30_1sigma} we plot the difference in radii of twin stars formed with the mass 0.77 $M_{\odot}$. One of these two is in the hybrid branch with lesser radii (marked in the earlier figure with stars) and another is in the hadron branch with radius =12.15km. It is observed that the difference in radii of these twin stars increases with the energy gap in similar fashion for all values of the speed of sound. This implies that with increase in energy gap, compact objects are formed with lesser radii in the hybrid branch leading to increase in the difference $\Delta R$.  The difference $\Delta R$ is most when sound speed is the least though the change is not that significant.

In the next figure Fig.~\ref{fig:mr_0.40}, we increase the transition density to 0.40 $\text{fm}^{-3}$.  The mass and  radii of the small object is  satisfied by the model calculations for the energy gap parameter ($\Delta \varepsilon$) 900 MeV $\text{fm}^{-3} $ and 1100 MeV $\text{fm}^{-3} $ at 1$\sigma$ level  and  that of 500 MeV $\text{fm}^{-3} $ at  2$\sigma$ level.   The results with the lowest energy gap satisfy the constraints from  GW170817. In the right panel of this figure, we plot the mass of those objects from our calculations which have radii equal to 10.4 km. The masses lie within the observed range for the range of  energy gap as is seen from the figure. Due to high transition density, $M=0.77 M_{\odot}$ is not achievable within the stable hybrid branch as is observed from the figure. Consequently, the radius corresponding to  $M=0.77 M_{\odot} $ is also not attainable.

\begin{figure*}[htp]
    \centering
      \includegraphics[width=0.45\textwidth]{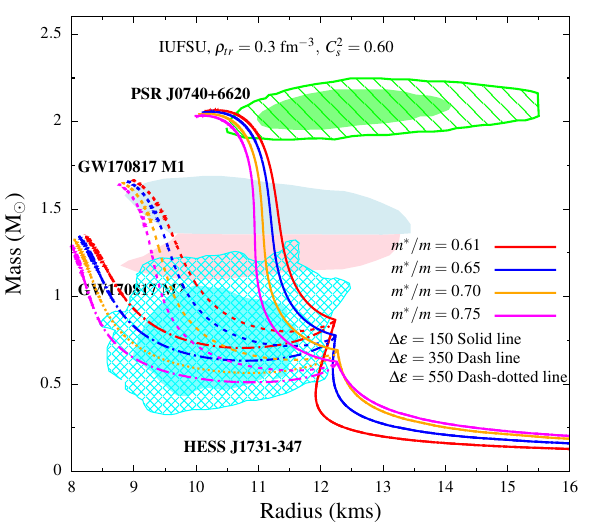}
       \includegraphics[width=0.45\textwidth]{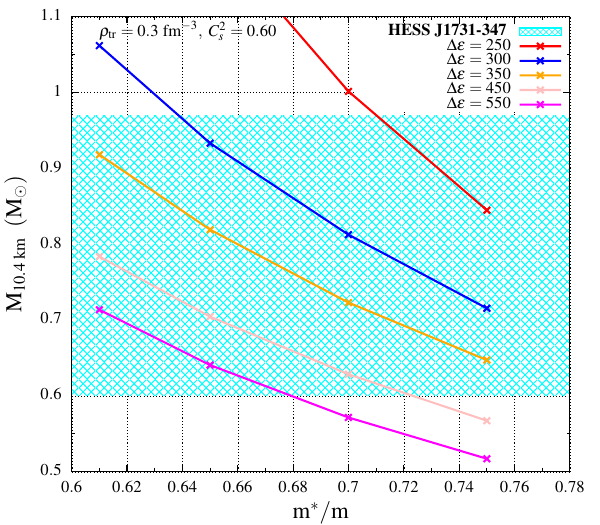}
        \includegraphics[width=0.45\textwidth]{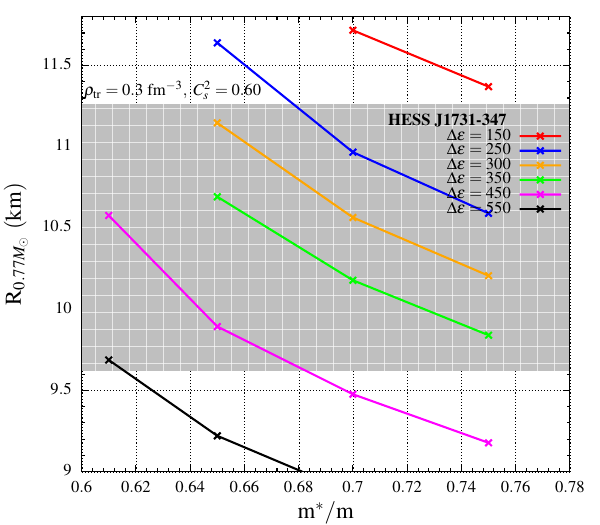}
\caption{\it  The top left panel shows the mass-radius relationship for the different values of $m^*/m$ along with different values of $\Delta \varepsilon$ for the transition density ($\rho_{\text{tr}}$)=0.3 $\text{fm}^{-3}$ and $C_s^2=0.6$.
The top right panel shows the variation of $M_{\text{10.4km}}$ with $m^*/m$ for different values of $\Delta \varepsilon$  with a fixed radius, $R_{1\sigma}^{\text{HESS}} = 10.4$ km.  The observed mass range ($M=0.77_{-0.17}^{+0.20}M_{\odot}$) has been shaded with cyan . The bottom panel illustrates the variation of $R_{0.77 M_{\odot}}$ with $m^*/m$ for different values of $\Delta \varepsilon$, fixing, $M_{1\sigma}^{\text{HESS}} = 0.77 M_{\odot}$.
The observed radii range ($R=10.4_{-0.78}^{+0.86} km$) has been shaded with grey. }
\label{fig:mr_mstar}
\end{figure*}

\begin{figure*}[htp]
    \centering
      \includegraphics[width=0.45\textwidth]{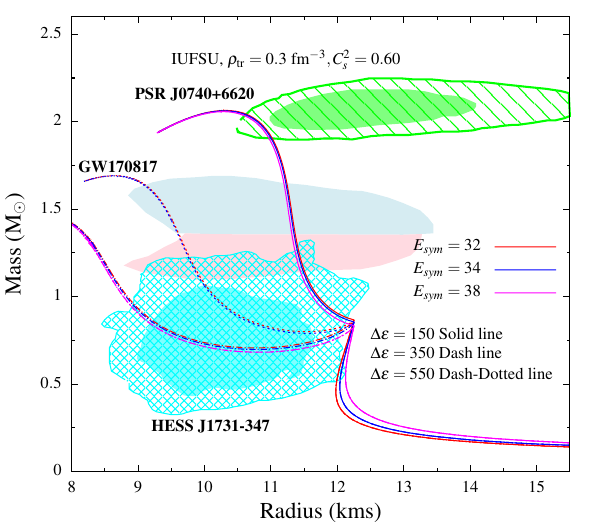}
       \includegraphics[width=0.45\textwidth]{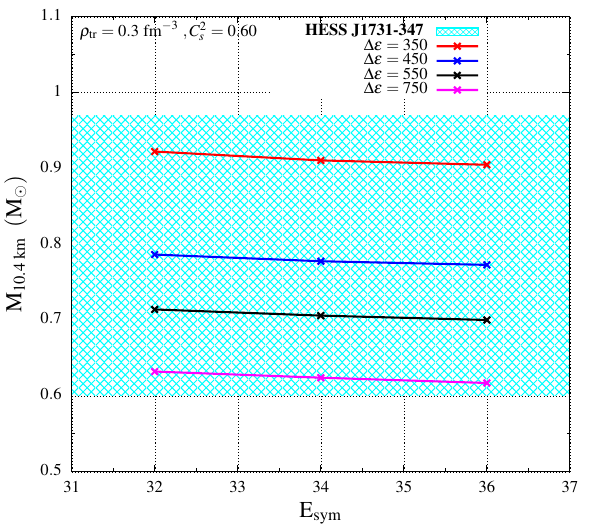}
        \includegraphics[width=0.45\textwidth]{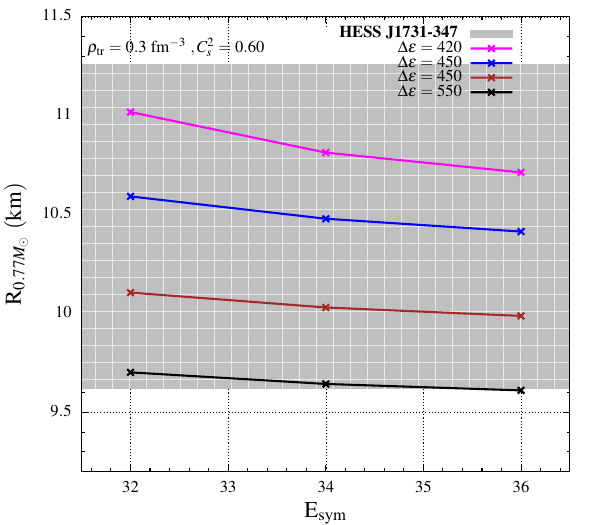}
\caption{\it  Same as Fig.6  for  different values of $E_{sym}$}
\label{fig:mr_Esym}
\end{figure*}

\begin{figure*}[htp]
    \centering
      \includegraphics[width=0.45\textwidth]{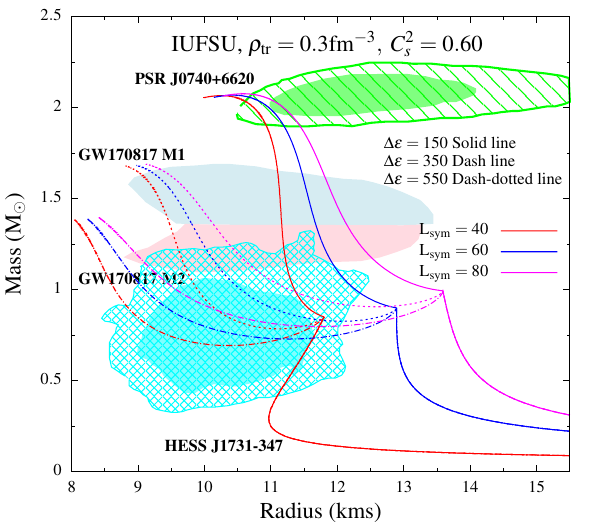}
      \includegraphics[width=0.45\textwidth]{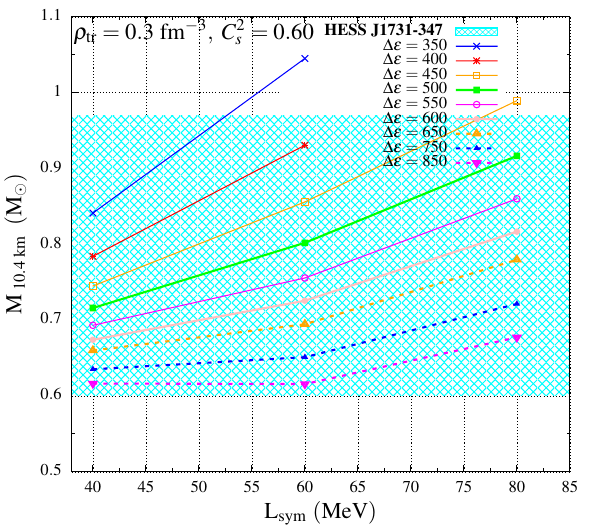}
        \includegraphics[width=0.45\textwidth]{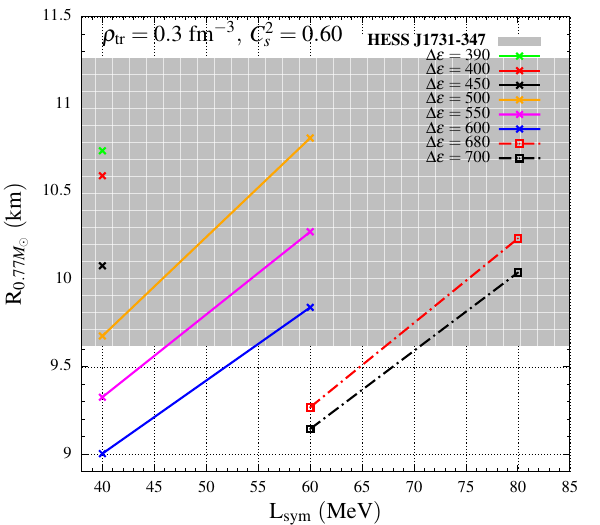}
\caption{\it   Same as Fig. 6  for  different values of $L_{sym}$}
\label{fig:mr_Lsym}
\end{figure*}

\section{Variation of the parameters of the hadronic equation of state}\label{sec:Variation of the parameters of the hadronic EoS}

Here we examine how hadronic parameters influence the structural properties of the hybrid stars. We have used the RMF model with  IUFSU parametrizations; the choice of the coupling constants depends on the nuclear matter properties at the saturation densities like effective mass  parameter $\text{m}^*/\text{m}$, symmetry energy$(\text{E}_{\text{sym}})$, the slope of the symmetry energy $(\text{L}_{\text{sym}})$ and others 
 as shown in  Table[\ref{tab:1}]. In this work, we take into account variation in the $\text{m}^*/\text{m}$, $\text{E}_{\text{sym}}$ and $\text{L}_{\text{sym}}$ parameters.

First, we have studied the effect of the effective mass parameter $\text{m}^*/\text{m}$ by varying it in the range of $0.61-0.75$. We have fixed the transition density at $\rho_{\text{tr}}=0.3 \, \text{fm}^{-3}$ and $C_s^2=0.6$, while varying $\Delta \varepsilon$ within the range of $150-550$ MeV $\text{fm}^{-3} $, as shown in Fig.~\ref{fig:mr_mstar}. In the left panel of the figure, we have displayed the mass-radius diagram.
We observe that changes in the effective mass parameter $(m^*/m)$ significantly affect the hybrid star configurations. For higher values of the effective mass parameter, the equation of state (EoS) becomes softer. The parameter $\text{m}^*/\text{m}$ also has a substantial influence on the maximum mass of the twin star configuration.
For lower values of $\Delta \varepsilon$ (e.g., $150~\text{MeV}${$\text{fm}^{-3} $}), the maximum mass constraints from PSR J0740+6620 are satisfied. As $\Delta \varepsilon$ increases, it enhances the compatibility with the HESS J1731-347 constraints and promotes the formation of the twin star configuration.
In Fig.~\ref{fig:mr_mstar}(right panel), we have plotted the masses ($M_{10.4km}$) of the compact objects formed in the stable hybrid branch from our sets of  EoS having radius =10.4 km by  varying $\text{m}^*/\text{m}$. Our analysis shows that as the value of $\text{m}^*/\text{m}$ increases, the corresponding values of $M_{10.04 km}$ decrease. For lower values of $\text{m}^*/\text{m}$, smaller values of $\Delta \varepsilon$ (e.g., 250 and 300 MeV $\text{fm}^{-3} $) fail to satisfy the constraints. Conversely, for higher values of $\text{m}^*/\text{m}$, larger values of $\Delta \varepsilon$ (e.g., 450 and 550 MeV $\text{fm}^{-3} $) are inconsistent with the observations. The most suitable value for $\Delta \varepsilon$ is found to be 350 MeV $\text{fm}^{-3} $ which satisfy the constraints on mass for the entire range of the effective mass parameter used. 
In Fig.~\ref{fig:mr_mstar}(bottom panel), we have plotted the radii $R_{0.77 M_{\odot}}$ obtained for different compact stars corresponding to the mass 0.77$M_{\odot}$ as estimated for the  compact object in the remnant of HESS J1731-347 . Our analysis shows that as the value of $m^*/m$ increases, the corresponding values of $\text{R}_{0.77 M_{\odot}}$ decrease. For lower values of $\text{m}^*/\text{m}$ (e.g., 0.61), $\Delta \varepsilon = 450$ and $550 \, \text{MeV}$ $\text{fm}^{-3} $ are allowed. In the intermediate range of $\text{m}^*/\text{m}$ (0.68–0.70), a broader range of $\Delta \varepsilon$ values is permitted. For higher values of $\text{m}^*/\text{m}$ (e.g., 0.75), the allowed values of $\Delta \varepsilon$ are restricted to approximately 250–350 MeV $\text{fm}^{-3} $.

Next, we have studied the effect of $E_{\text{sym}}$ by varying it in the range of 32–36 MeV. We  have fixed the transition density at $\rho_{\text{tr}}=0.3 \, \text{fm}^{-3}$ and $C_s^2=0.6$, while varying $\Delta \varepsilon$ within the range of 150–550 MeV $\text{fm}^{-3} $, as shown in Fig.~\ref{fig:mr_Esym}. In Fig.~\ref{fig:mr_Esym}(left panel), we present the mass-radius diagram. Our findings indicate that the impact of $\text{E}_{\text{sym}}$ on the structural properties of the hybrid star is relatively weak and not significantly influential.
In Fig.~\ref{fig:mr_Esym}(  right panel ), we  have shown constrained the mass of  the hybrid stars ($M_{10.4km}$)  by varying $\text{E}_{\text{sym}}$, with the radius fixed at $\text{R}_{\text{HESS}}=10.4 \, \text{km}$. Our analysis shows that the value of $\text{M}_{\text{10.4 km}}$ remains almost constant with variations in $E_{sym}$. The allowed range for $\Delta \varepsilon$ is approximately  350–750 MeV $\text{fm}^{-3} $.
In Fig.~\ref{fig:mr_Esym}(bottom panel), we have plotted the radius  
 $R_{0.77 M_{\odot}}$ by  varying $E_{sym}$, with the mass fixed at $0.77M_{\odot}$. Our analysis shows that the value of $R_{0.77 M_{\odot}}$ varies only slightly with changes in $E_{sym}$. The allowed range for $\Delta \varepsilon$ is nearly about 420–550 MeV $\text{fm}^{-3} $.
In the case of $E_{sym}$ variation, the allowed range of $\Delta \varepsilon$  is more restricted for $R_{0.77 M_{\odot}}$ (about 130 MeV $\text{fm}^{-3} $) as compared to the case of $M_{10.4 km}$ (nearly 400 MeV $\text{fm}^{-3} $) as seen from the figures ~\ref{fig:mr_Esym} right panel and middle panel.

We then investigate the effect of the variation of the slope of the symmetry energy parameter $L_{\text{sym}}$   keeping  the other parameters unchanged. In Fig.~\ref{fig:mr_Lsym} we illustrate the effect of $L_{\text{sym}}$ on the mass-radius, for $\rho_{\text{tr}}=0.3 ~\text{fm}^{-3}$. We have found that the effect $L_{\text{sym}}$ is much more pronounced as compared to $E_{\text{sym}}$, significantly affecting the hybrid star configurations, specially the radii of the compact objects. The radii on the hadron branch is more for higher values of $L_{\text{sym}}$. For each $L_{\text{sym}}$, we have varied the energy gap parameter. From  Fig.~\ref{fig:mr_Lsym}( left panel ),  it has been observed that for  the lowest value of $L_{sym}$ (40 MeV), the results are compatible with the compact object in HESS J1731-347 remnant for the chosen values of the energy gap (150 MeV $\text{fm}^{-3} $-550 MeV $\text{fm}^{-3} $). The compact objects in the hadron branch as well as those after the phase transition satisfy the constraints from this  compact object.  As $L_{\text{sym}}$ is increased, lower values of the energy gap goes outside the domain of mass and radius of  this compact object.  The constraint of 2$M_{\odot}$ from PSR J0740+6620 is satisfied by the energy gap of 150 MeV $\text{fm}^{-3} $ for all chosen values of $L_{\text{sym}}$ (40-80 MeV). 
Next,we have plotted the masses of the stars ($M_{10.4km}$) corresponding to the radii of 10.4 km as a function of $L_{\text{sym}}$ for different energy density gap parameters in Fig.~\ref{fig:mr_Lsym} (right panel). The masses increase with $L_{\text{sym}}$ and for a fixed $L_{\text{sym}}$, they decrease with the gap parameter. For the lowest value of $L_{\text{sym}}$ (40 MeV), the masses lie well within the observed range but for the higher values, few are outside the range for lower energy gaps. 
In Fig.~\ref{fig:mr_Lsym} (bottom panel ), we have plotted the radii $R_{0.77 M_{\odot}}$ corresponding to mass of 0.77 $M_{\odot}$ and it is  being observed that  lower energy gaps satisfy the constraints for lower $L_{\text{sym}}$ and as  $L_{\text{sym}}$ is increased, higher energy gaps are required to satisfy the constraint of radii. This is in contrast with the previous diagram (middle panel) where almost the entire range of energy gap satisfies the constraint of mass for all the values of $L_{\text{sym}}$.

We have changed  the three parameters of the hadronic equations of state ,  and in each case we have found that for the lower values of the energy density gap, configuration reaches the maximum mass exceeding 2$M_{\odot}$ (left panel of Fig.~\ref{fig:mr_mstar}, \ref{fig:mr_Esym} and \ref{fig:mr_Lsym} ) and at the same time it also satisfies the HESS J1731-347  data at 2$\sigma$ limits.
\color{black}

\section{  SUMMARY AND FUTURE OUTLOOK}\label{sec:conclusion}

 The main goal of our work is to explain the nature of the CCO HESS J1731-347 considering it to be a hybrid star using the constant speed of sound parameters. The EoS is based on the IUFSU and CSS parametrizations, and the phase transition between hadronic and quark matter is modeled via  Maxwell construction (MC). The mass-radius relationships of these stars are examined by varying parameters such as the transition density, energy gap,  and speed of sound in the quark matter sector. For different values of $\rho_{tr}$, the EoS is constrained by observed data, including the mass-radius observations of compact objects such as PSR J0740+6620, HESS J1731-347, and the GW170817 event. 
For a lucid visualization of the parameter space, we fix the radius of HESS J1731-347 at $10.4 \, \text{km}$ and constrain the parameters for the allowed mass range, $M_{\text{HESS}} = 0.77^{+0.20}_{-0.17} \, M_{\odot}$, for a given transition density. Similarly, by fixing the mass at $ 0.77 \, M_{\odot}$, we constrain the parameter space for the allowed radius range, $R_{\text{HESS}} = 10.4^{+0.86}_{-0.78} \, \text{km}$. We have observed that the effect of  the gap parameter $\Delta \varepsilon$ is significantly more pronounced than that of $C_s^2$.
HESS J1731-347  also provides a unique opportunity to explore low-mass twin stars. We investigate the possible radius difference of the twin stars with a mass of $M_{\text{twin}} = 0.77 \, M_{\odot}$. Our findings reveal that $\Delta R_{\text{HESS}}$ increases with $\Delta \varepsilon$ for a fixed $C_s^2$, while it decreases with $C_s^2$ for a fixed $\Delta \varepsilon$. We have also studied the effect the hadronic parameters $m^*/m$, $E_{sym} $ and $L_{sym}$. It has been observed that $m*/m$ and $L_{sym}$ have a pronounced effect on mass and radii of the  compact object  but the influence of  $E_{sym}$ is not that significant.
Our findings align with the hypothesis that the compact object in HESS J1731-347 is likely a stable hybrid star with quark matter in the inner core.

Our calculation and analysis reveal that the observational data obtained from  HESS J1731-347 throw light in
the range of 1-2 nuclear saturation density in contrast to GW170817, GW190425,
NICER, and  other high mass  compact star measurements which  probe the
strongly interacting matter at higher densities. Further observation of such objects is necessary to have better idea on the constituents as well as about the existence of the low mass twin stars.

The study of first-order phase transitions (FOPTs) in neutron stars holds profound implications for future observations, particularly with the advancements in observational technologies and theoretical modeling. Next-generation detectors like the Einstein Telescope and Cosmic Explorer will enable more precise measurements of tidal deformability and post-merger oscillations. Improved sensitivity in the kilohertz frequency range will allow for more accurate measurements of post-merger oscillations, which could provide insights into the softening of the EoS and the possible formation of a quark core. X-ray missions such as Athena and STROBE-X will enhance mass-radius constraints which may help distinguish between hadronic and hybrid EoS. The Square Kilometer Array (SKA) will significantly enhance measurements of neutron star masses, especially for ultra-massive stars above 2.5$M_{\odot}$, which are more likely to support EoS with phase transitions. Additionally, precise pulsar timing will aid in detecting glitch behavior and changes in the moment of inertia, offering valuable insights into the evolution of the star's core structure. By combining gravitational wave and X-ray data through Bayesian inference, one can refine constraints on the EoS. On the other hand, general relativistic hydrodynamics simulations that incorporate advanced microphysics, such as neutrino transport and magnetic field dynamics, will provide more accurate predictions for gravitational wave signals associated with phase transitions. The search for twin-star configurations, where stars of similar masses exhibit different radii, could provide direct evidence for phase transitions. Upcoming mass-radius data from X-ray and gravitational wave observations could validate this. Together, these advancements will be crucial for unraveling the nature of dense matter and the role of FOPTs in neutron star structure and evolution.

\color{black}
\bibliography{hess_hybrid.bib}

\begin{thebibliography}{}
\expandafter\ifx\csname natexlab\endcsname\relax\def\natexlab#1{#1}\fi
\providecommand{\url}[1]{\href{#1}{#1}}
\providecommand{\dodoi}[1]{doi:~\href{http://doi.org/#1}{\nolinkurl{#1}}}
\providecommand{\doeprint}[1]{\href{http://ascl.net/#1}{\nolinkurl{http://ascl.net/#1}}}
\providecommand{\doarXiv}[1]{\href{https://arxiv.org/abs/#1}{\nolinkurl{https://arxiv.org/abs/#1}}}

\bibitem[{Abbott {et~al.}(2018)}]{LIGOScientific:2018cki}
Abbott, B.~P., {et~al.} 2018, Phys. Rev. Lett., 121, 161101,
  \dodoi{10.1103/PhysRevLett.121.161101}

\bibitem[{Alford {et~al.}(2013)Alford, Han, \& Prakash}]{PhysRevD.88.083013}
Alford, M.~G., Han, S., \& Prakash, M. 2013, Phys. Rev. D, 88, 083013,
  \dodoi{10.1103/PhysRevD.88.083013}

\bibitem[{Altiparmak {et~al.}(2022)Altiparmak, Ecker, \&
  Rezzolla}]{Altiparmak_2022}
Altiparmak, S., Ecker, C., \& Rezzolla, L. 2022, The Astrophysical Journal
  Letters, 939, L34, \dodoi{10.3847/2041-8213/ac9b2a}

\bibitem[{{Baym} {et~al.}(1971){Baym}, {Pethick}, \& {Sutherland}}]{BPS_1971}
{Baym}, G., {Pethick}, C., \& {Sutherland}, P. 1971, \apj, 170, 299,
  \dodoi{10.1086/151216}

\bibitem[{Benvenuto \& Lugones(1995)}]{Benvenuto95}
Benvenuto, O.~G., \& Lugones, G. 1995, Phys. Rev. D, 51, 1989,
  \dodoi{10.1103/PhysRevD.51.1989}

\bibitem[{Borsányi {et~al.}(2014)Borsányi, Fodor, Hoelbling, Katz, Krieg, \&
  Szabó}]{BORSANYI201499}
Borsányi, S., Fodor, Z., Hoelbling, C., {et~al.} 2014, Physics Letters B, 730,
  99, \dodoi{https://doi.org/10.1016/j.physletb.2014.01.007}

\bibitem[{Buballa(2005)}]{Buballa:2003qv}
Buballa, M. 2005, Phys. Rept., 407, 205, \dodoi{10.1016/j.physrep.2004.11.004}

\bibitem[{Buballa \& Oertel(1999)}]{Buballa:1998pr}
Buballa, M., \& Oertel, M. 1999, Phys. Lett. B, 457, 261,
  \dodoi{10.1016/S0370-2693(99)00533-X}

\bibitem[{Carriere {et~al.}(2003)Carriere, Horowitz, \&
  Piekarewicz}]{Carriere:2002bx}
Carriere, J., Horowitz, C.~J., \& Piekarewicz, J. 2003, Astrophys. J., 593,
  463, \dodoi{10.1086/376515}

\bibitem[{Char \& Biswas(2024)}]{Char:2024kgo}
Char, P., \& Biswas, B. 2024.
\newblock \doarXiv{2408.15220}

\bibitem[{Chen \& Piekarewicz(2014)}]{Chen:2014sca}
Chen, W.-C., \& Piekarewicz, J. 2014, Phys. Rev. C, 90, 044305,
  \dodoi{10.1103/PhysRevC.90.044305}

\bibitem[{Cherman {et~al.}(2009)Cherman, Cohen, \&
  Nellore}]{PhysRevD.80.066003}
Cherman, A., Cohen, T.~D., \& Nellore, A. 2009, Phys. Rev. D, 80, 066003,
  \dodoi{10.1103/PhysRevD.80.066003}

\bibitem[{Chodos {et~al.}(1974)Chodos, Jaffe, Johnson, Thorn, \&
  Weisskopf}]{chodes1974}
Chodos, A., Jaffe, R.~L., Johnson, K., Thorn, C.~B., \& Weisskopf, V.~F. 1974,
  Phys. Rev. D, 9, 3471, \dodoi{10.1103/PhysRevD.9.3471}

\bibitem[{Choudhury {et~al.}(2024)}]{Choudhury:2024xbk}
Choudhury, D., {et~al.} 2024, Astrophys. J. Lett., 971, L20,
  \dodoi{10.3847/2041-8213/ad5a6f}

\bibitem[{Chu \& Chen(2013)}]{Chu_2014}
Chu, P.-C., \& Chen, L.-W. 2013, The Astrophysical Journal, 780, 135,
  \dodoi{10.1088/0004-637X/780/2/135}

\bibitem[{Chu {et~al.}(2023)Chu, Li, Liu, Ju, \& Zhou}]{pen:23prc_qmdd}
Chu, P.-C., Li, X.-H., Liu, H., Ju, M., \& Zhou, Y. 2023, Phys. Rev. C, 108,
  025808, \dodoi{10.1103/PhysRevC.108.025808}

\bibitem[{Dittmann {et~al.}(2024)}]{Dittmann:2024mbo}
Dittmann, A.~J., {et~al.} 2024, Astrophys. J., 974, 295,
  \dodoi{10.3847/1538-4357/ad5f1e}

\bibitem[{{Doroshenko} {et~al.}(2022){Doroshenko}, {Suleimanov},
  {P{\"u}hlhofer}, \& {Santangelo}}]{2022NatAs}
{Doroshenko}, V., {Suleimanov}, V., {P{\"u}hlhofer}, G., \& {Santangelo}, A.
  2022, Nature Astronomy, 6, 1444, \dodoi{10.1038/s41550-022-01800-1}

\bibitem[{et~all(2021)}]{Adhikari_PRL_2021}
et~all, A. 2021, Phys. Rev. Lett., 126, 172502,
  \dodoi{10.1103/PhysRevLett.126.172502}

\bibitem[{Fattoyev {et~al.}(2010)Fattoyev, Horowitz, Piekarewicz, \&
  Shen}]{Piekarewicz_prc_2010_nov}
Fattoyev, F.~J., Horowitz, C.~J., Piekarewicz, J., \& Shen, G. 2010, Phys. Rev.
  C, 82, 055803, \dodoi{10.1103/PhysRevC.82.055803}

\bibitem[{Fonseca {et~al.}(2021)}]{Fonseca:2021wxt}
Fonseca, E., {et~al.} 2021, Astrophys. J. Lett., 915, L12,
  \dodoi{10.3847/2041-8213/ac03b8}

\bibitem[{Fraga {et~al.}(2001)Fraga, Pisarski, \&
  Schaffner-Bielich}]{Fraga2001}
Fraga, E.~S., Pisarski, R.~D., \& Schaffner-Bielich, J. 2001, Phys. Rev. D, 63,
  121702, \dodoi{10.1103/PhysRevD.63.121702}

\bibitem[{Gholami {et~al.}(2024)Gholami, Rather, Hofmann, Buballa, \&
  Schaffner-Bielich}]{Gholami:2024ety}
Gholami, H., Rather, I.~A., Hofmann, M., Buballa, M., \& Schaffner-Bielich, J.
  2024.
\newblock \doarXiv{2411.04064}

\bibitem[{Glendenning(2000)}]{Glendenning:1997wn}
Glendenning, N.~K. 2000, {Compact stars: Nuclear physics, particle physics, and
  general relativity} (Springer-Verlag, New York)

\bibitem[{Han {et~al.}(2019)Han, Mamun, Lalit, Constantinou, \&
  Prakash}]{Han:2019bub}
Han, S., Mamun, M. A.~A., Lalit, S., Constantinou, C., \& Prakash, M. 2019,
  Phys. Rev. D, 100, 103022, \dodoi{10.1103/PhysRevD.100.103022}

\bibitem[{Hanauske {et~al.}(2001)Hanauske, Satarov, Mishustin, Stoecker, \&
  Greiner}]{Hanauske:2001nc}
Hanauske, M., Satarov, L.~M., Mishustin, I.~N., Stoecker, H., \& Greiner, W.
  2001, Phys. Rev. D, 64, 043005, \dodoi{10.1103/PhysRevD.64.043005}

\bibitem[{Hatsuda \& Kunihiro(1994)}]{Hatsuda:1994pi}
Hatsuda, T., \& Kunihiro, T. 1994, Phys. Rept., 247, 221,
  \dodoi{10.1016/0370-1573(94)90022-1}

\bibitem[{Hong \& Ren(2024)}]{Hong:2024sey}
Hong, B., \& Ren, Z. 2024, Phys. Rev. D, 109, 023002,
  \dodoi{10.1103/PhysRevD.109.023002}

\bibitem[{Hornick {et~al.}(2018)Hornick, Tolos, Zacchi, Christian, \&
  Schaffner-Bielich}]{Hornick:2018kfi}
Hornick, N., Tolos, L., Zacchi, A., Christian, J.-E., \& Schaffner-Bielich, J.
  2018, Phys. Rev. C, 98, 065804, \dodoi{10.1103/PhysRevC.98.065804}

\bibitem[{Horvath {et~al.}(2023)Horvath, Rocha, de~S\'a, Moraes, Bar\~ao,
  de~Avellar, Bernardo, \& Bachega}]{Horvath:2023uwl}
Horvath, J.~E., Rocha, L.~S., de~S\'a, L.~M., {et~al.} 2023, Astron.
  Astrophys., 672, L11, \dodoi{10.1051/0004-6361/202345885}

\bibitem[{Kanakis-Pegios {et~al.}(2020)Kanakis-Pegios, Koliogiannis, \&
  Moustakidis}]{PhysRevC.102.055801}
Kanakis-Pegios, A., Koliogiannis, P.~S., \& Moustakidis, C.~C. 2020, Phys. Rev.
  C, 102, 055801, \dodoi{10.1103/PhysRevC.102.055801}

\bibitem[{Klevansky(1992)}]{Klevansky92}
Klevansky, S.~P. 1992, Rev. Mod. Phys., 64, 649,
  \dodoi{10.1103/RevModPhys.64.649}

\bibitem[{Kourmpetis {et~al.}(2024)Kourmpetis, Laskos-Patkos, \&
  Moustakidis}]{Kourmpetis:2024mol}
Kourmpetis, K., Laskos-Patkos, P., \& Moustakidis, C.~C. 2024.
\newblock \doarXiv{2411.17234}

\bibitem[{Kubis {et~al.}(2023)Kubis, W\'ojcik, Castillo, \&
  Zabari}]{Kubis:2023gxa}
Kubis, S., W\'ojcik, W., Castillo, D.~A., \& Zabari, N. 2023, Phys. Rev. C,
  108, 045803, \dodoi{10.1103/PhysRevC.108.045803}

\bibitem[{Laskos-Patkos {et~al.}(2024)Laskos-Patkos, Koliogiannis, \&
  Moustakidis}]{Laskos-Patkos:2023tlr}
Laskos-Patkos, P., Koliogiannis, P.~S., \& Moustakidis, C.~C. 2024, Phys. Rev.
  D, 109, 063017, \dodoi{10.1103/PhysRevD.109.063017}

\bibitem[{Laskos-Patkos \& Moustakidis(2023)}]{Laskos-Patkos:2023cts}
Laskos-Patkos, P., \& Moustakidis, C.~C. 2023, Phys. Rev. D, 107, 123023,
  \dodoi{10.1103/PhysRevD.107.123023}

\bibitem[{Lenzi {et~al.}(2010)Lenzi, Schneider, Providencia, \&
  Marinho}]{Lenzi:2010mz}
Lenzi, C.~H., Schneider, A.~S., Providencia, C., \& Marinho, R.~M. 2010, Phys.
  Rev. C, 82, 015809, \dodoi{10.1103/PhysRevC.82.015809}

\bibitem[{Ma {et~al.}(2023)Ma, Lu, Xu, Peng, Fu, \& Wang}]{Ma:prd23sep}
Ma, Z.-J., Lu, Z.-Y., Xu, J.-F., {et~al.} 2023, Phys. Rev. D, 108, 054017,
  \dodoi{10.1103/PhysRevD.108.054017}

\bibitem[{Margaritis {et~al.}(2020)Margaritis, Koliogiannis, \&
  Moustakidis}]{Margaritis:2019hfq}
Margaritis, C., Koliogiannis, P.~S., \& Moustakidis, C.~C. 2020, Phys. Rev. D,
  101, 043023, \dodoi{10.1103/PhysRevD.101.043023}

\bibitem[{Mariani {et~al.}(2024)Mariani, Ranea-Sandoval, Lugones, \&
  Orsaria}]{Mariani:2024gqi}
Mariani, M., Ranea-Sandoval, I.~F., Lugones, G., \& Orsaria, M.~G. 2024, Phys.
  Rev. D, 110, 043026, \dodoi{10.1103/PhysRevD.110.043026}

\bibitem[{Maruyama {et~al.}(2007)Maruyama, Chiba, Schulze, \&
  Tatsumi}]{Maruyama:2007ey}
Maruyama, T., Chiba, S., Schulze, H.-J., \& Tatsumi, T. 2007, Phys. Rev. D, 76,
  123015, \dodoi{10.1103/PhysRevD.76.123015}

\bibitem[{Maruyama {et~al.}(2008)Maruyama, Chiba, Schulze, \&
  Tatsumi}]{Maruyama:2007ss}
---. 2008, Phys. Lett. B, 659, 192, \dodoi{10.1016/j.physletb.2007.10.056}

\bibitem[{Miller {et~al.}(2019)}]{Miller:2019cac}
Miller, M.~C., {et~al.} 2019, Astrophys. J. Lett., 887, L24,
  \dodoi{10.3847/2041-8213/ab50c5}

\bibitem[{Miller {et~al.}(2021)}]{Miller:2021qha}
---. 2021, Astrophys. J. Lett., 918, L28, \dodoi{10.3847/2041-8213/ac089b}

\bibitem[{Monta\~na {et~al.}(2019)Monta\~na, Tol\'os, Hanauske, \&
  Rezzolla}]{prd_Monta_2019_may}
Monta\~na, G., Tol\'os, L., Hanauske, M., \& Rezzolla, L. 2019, Phys. Rev. D,
  99, 103009, \dodoi{10.1103/PhysRevD.99.103009}

\bibitem[{Montana {et~al.}(2019)Montana, Tolos, Hanauske, \&
  Rezzolla}]{Montana:2018bkb}
Montana, G., Tolos, L., Hanauske, M., \& Rezzolla, L. 2019, Phys. Rev. D, 99,
  103009, \dodoi{10.1103/PhysRevD.99.103009}

\bibitem[{Nambu \& Jona-Lasinio(1961)}]{nambu1961njl}
Nambu, Y., \& Jona-Lasinio, G. 1961, Phys. Rev., 122, 345,
  \dodoi{10.1103/PhysRev.122.345}

\bibitem[{Oppenheimer \& Volkoff(1939)}]{PhysRev.55.374}
Oppenheimer, J.~R., \& Volkoff, G.~M. 1939, Phys. Rev., 55, 374,
  \dodoi{10.1103/PhysRev.55.374}

\bibitem[{Pal \& Chaudhuri(2023)}]{Pal:2023dlv}
Pal, S., \& Chaudhuri, G. 2023, Phys. Rev. D, 108, 103028,
  \dodoi{10.1103/PhysRevD.108.103028}

\bibitem[{Pal \& Chaudhuri(2024{\natexlab{a}})}]{Pal:2024afl}
---. 2024{\natexlab{a}}, JCAP, 10, 064, \dodoi{10.1088/1475-7516/2024/10/064}

\bibitem[{Pal \& Chaudhuri(2024{\natexlab{b}})}]{Pal:2024nza}
---. 2024{\natexlab{b}}, Phys. Rev. D, 110, 123021,
  \dodoi{10.1103/PhysRevD.110.123021}

\bibitem[{Pal {et~al.}(2023{\natexlab{a}})Pal, Podder, Sen, \&
  Chaudhuri}]{suman_2023_prd1}
Pal, S., Podder, S., Sen, D., \& Chaudhuri, G. 2023{\natexlab{a}}, Phys. Rev.
  D, 107, 063019, \dodoi{10.1103/PhysRevD.107.063019}

\bibitem[{Pal {et~al.}(2023{\natexlab{b}})Pal, Podder, Sen, \&
  Chaudhuri}]{Pal:2023quk}
---. 2023{\natexlab{b}}, Phys. Rev. D, 107, 063019,
  \dodoi{10.1103/PhysRevD.107.063019}

\bibitem[{Peng {et~al.}(1999)Peng, Chiang, Yang, Li, \& Liu}]{peng2001a}
Peng, G.~X., Chiang, H.~C., Yang, J.~J., Li, L., \& Liu, B. 1999, Phys. Rev. C,
  61, 015201, \dodoi{10.1103/PhysRevC.61.015201}

\bibitem[{Pereira {et~al.}(2022)Pereira, Bejger, Zdunik, \&
  Haensel}]{Pereira:2022stw}
Pereira, J.~P., Bejger, M., Zdunik, J.~L., \& Haensel, P. 2022, Phys. Rev. D,
  105, 123015, \dodoi{10.1103/PhysRevD.105.123015}

\bibitem[{Pfaff {et~al.}(2022)Pfaff, Hansen, \& Gulminelli}]{Pfaff:2021kse}
Pfaff, A., Hansen, H., \& Gulminelli, F. 2022, Phys. Rev. C, 105, 035802,
  \dodoi{10.1103/PhysRevC.105.035802}

\bibitem[{Plumari {et~al.}(2013)Plumari, Burgio, Greco, \&
  Zappal\`a}]{Plumari2013}
Plumari, S., Burgio, G.~F., Greco, V., \& Zappal\`a, D. 2013, Phys. Rev. D, 88,
  083005, \dodoi{10.1103/PhysRevD.88.083005}

\bibitem[{Podder {et~al.}(2024)Podder, Pal, Sen, \& Chaudhuri}]{Podder:2023dey}
Podder, S., Pal, S., Sen, D., \& Chaudhuri, G. 2024, Nucl. Phys. A, 1042,
  122796, \dodoi{10.1016/j.nuclphysa.2023.122796}

\bibitem[{Ranea-Sandoval {et~al.}(2016)Ranea-Sandoval, Han, Orsaria, Contrera,
  Weber, \& Alford}]{Ranea-Sandoval:2015ldr}
Ranea-Sandoval, I.~F., Han, S., Orsaria, M.~G., {et~al.} 2016, Phys. Rev. C,
  93, 045812, \dodoi{10.1103/PhysRevC.93.045812}

\bibitem[{Rather {et~al.}(2023)Rather, Panotopoulos, \& Lopes}]{Rather:2023tly}
Rather, I.~A., Panotopoulos, G., \& Lopes, I. 2023, Eur. Phys. J. C, 83, 1065,
  \dodoi{10.1140/epjc/s10052-023-12223-1}

\bibitem[{Reardon {et~al.}(2024)}]{Reardon:2024rdv}
Reardon, D.~J., {et~al.} 2024, Astrophys. J. Lett., 971, L18,
  \dodoi{10.3847/2041-8213/ad614a}

\bibitem[{Riley {et~al.}(2019)}]{Riley:2019yda}
Riley, T.~E., {et~al.} 2019, Astrophys. J. Lett., 887, L21,
  \dodoi{10.3847/2041-8213/ab481c}

\bibitem[{Riley {et~al.}(2021)}]{Riley:2021pdl}
---. 2021, Astrophys. J. Lett., 918, L27, \dodoi{10.3847/2041-8213/ac0a81}

\bibitem[{Sagun {et~al.}(2023)Sagun, Giangrandi, Dietrich, Ivanytskyi,
  Negreiros, \& Provid\^encia}]{Sagun:2023rzp}
Sagun, V., Giangrandi, E., Dietrich, T., {et~al.} 2023, Astrophys. J., 958, 49,
  \dodoi{10.3847/1538-4357/acfc9e}

\bibitem[{Salmi {et~al.}(2022)}]{Salmi:2022cgy}
Salmi, T., {et~al.} 2022, Astrophys. J., 941, 150,
  \dodoi{10.3847/1538-4357/ac983d}

\bibitem[{Salmi {et~al.}(2024)}]{Salmi:2024aum}
---. 2024, Astrophys. J., 974, 294, \dodoi{10.3847/1538-4357/ad5f1f}

\bibitem[{Schertler {et~al.}(2000)Schertler, Greiner, Schaffner-Bielich, \&
  Thoma}]{SCHERTLER2000463}
Schertler, K., Greiner, C., Schaffner-Bielich, J., \& Thoma, M. 2000, Nuclear
  Physics A, 677, 463, \dodoi{https://doi.org/10.1016/S0375-9474(00)00305-5}

\bibitem[{Seidov(1971)}]{seidov1971stability}
Seidov, Z. 1971, Soviet Astronomy, Vol. 15, p. 347, 15, 347

\bibitem[{Sen {et~al.}(2021)Sen, Alam, \& Chaudhuri}]{Sen:2021cgl}
Sen, D., Alam, N., \& Chaudhuri, G. 2021, J. Phys. G, 48, 105201,
  \dodoi{10.1088/1361-6471/ac1713}

\bibitem[{Sen {et~al.}(2022)Sen, Alam, \& Chaudhuri}]{Sen:2022lig}
---. 2022, Phys. Rev. D, 106, 083008, \dodoi{10.1103/PhysRevD.106.083008}

\bibitem[{Sun \& Wen(2023)}]{Sun:2023glq}
Sun, H., \& Wen, D. 2023, Phys. Rev. C, 108, 025801,
  \dodoi{10.1103/PhysRevC.108.025801}

\bibitem[{Tan {et~al.}(2022)Tan, Dore, Dexheimer, Noronha-Hostler, \&
  Yunes}]{PhysRevD.105.023018}
Tan, H., Dore, T., Dexheimer, V., Noronha-Hostler, J., \& Yunes, N. 2022, Phys.
  Rev. D, 105, 023018, \dodoi{10.1103/PhysRevD.105.023018}

\bibitem[{Tewari {et~al.}(2024)Tewari, Chatterjee, Kumar, \&
  Mallick}]{Tewari:2024qit}
Tewari, S., Chatterjee, S., Kumar, D., \& Mallick, R. 2024.
\newblock \doarXiv{2410.20355}

\bibitem[{Tews {et~al.}(2018)Tews, Carlson, Gandolfi, \& Reddy}]{Tews:2018kmu}
Tews, I., Carlson, J., Gandolfi, S., \& Reddy, S. 2018, Astrophys. J., 860,
  149, \dodoi{10.3847/1538-4357/aac267}

\bibitem[{Tolman(1939)}]{PhysRev.55.364}
Tolman, R.~C. 1939, Phys. Rev., 55, 364, \dodoi{10.1103/PhysRev.55.364}

\bibitem[{Tsaloukidis {et~al.}(2023)Tsaloukidis, Koliogiannis, Kanakis-Pegios,
  \& Moustakidis}]{Tsaloukidis:2022rus}
Tsaloukidis, L., Koliogiannis, P.~S., Kanakis-Pegios, A., \& Moustakidis, C.~C.
  2023, Phys. Rev. D, 107, 023012, \dodoi{10.1103/PhysRevD.107.023012}

\bibitem[{Veselsky {et~al.}(2025)Veselsky, Koliogiannis, Petousis, Leja, \&
  Moustakidis}]{Veselsky:2024eae}
Veselsky, M., Koliogiannis, P.~S., Petousis, V., Leja, J., \& Moustakidis,
  C.~C. 2025, Phys. Lett. B, 860, 139185,
  \dodoi{10.1016/j.physletb.2024.139185}

\bibitem[{Wang {et~al.}(2020)Wang, Zhao, \& Zong}]{Wang:2020wzs}
Wang, Q., Zhao, T., \& Zong, H. 2020, Mod. Phys. Lett. A, 35, 2050321,
  \dodoi{10.1142/S0217732320503216}

\bibitem[{Wen {et~al.}(2005)Wen, Zhong, Peng, Shen, \& Ning}]{wen2005a}
Wen, X.~J., Zhong, X.~H., Peng, G.~X., Shen, P.~N., \& Ning, P.~Z. 2005, Phys.
  Rev. C, 72, 015204, \dodoi{10.1103/PhysRevC.72.015204}

\bibitem[{Xu(2003)}]{Xu2003}
Xu, R.-X. 2003, Astrophys. J. Lett., 596, L59, \dodoi{10.1086/379209}

\bibitem[{Zel'dovich(1961)}]{Zeldovich:1961sbr}
Zel'dovich, Y.~B. 1961, Zh. Eksp. Teor. Fiz., 41, 1609

\bibitem[{Zel’dovich \& Novikov(2014)}]{zel2014stars}
Zel’dovich, Y.~B., \& Novikov, I.~D. 2014, Stars and relativity (Courier
  Corporation)

\bibitem[{Zhang {et~al.}(2025)Zhang, Rueda~Hernandez, \&
  Negreiros}]{Zhang:2024ldq}
Zhang, S.-R., Rueda~Hernandez, J.~A., \& Negreiros, R. 2025, Astrophys. J.,
  978, 1, \dodoi{10.3847/1538-4357/ad96b5}

\bibitem[{Zhang {et~al.}(2021)Zhang, Chu, Li, Liu, \& Zhang}]{Zhang:2021qhl}
Zhang, Z., Chu, P.-C., Li, X.-H., Liu, H., \& Zhang, X.-M. 2021, Phys. Rev. D,
  103, 103021, \dodoi{10.1103/PhysRevD.103.103021}

\end{thebibliography}

\bibliographystyle{aasjournal}

\end{document}